\documentclass{article}[11pt]
\usepackage{graphicx}
\usepackage{microtype}
\usepackage{graphicx}
\usepackage{amsmath}
\usepackage{amsthm}
\usepackage{amsfonts}
\usepackage{amssymb}
\usepackage{clrscode3e}
\usepackage{bm}
\usepackage{upgreek}
\usepackage{tikz}
\usepackage{tikzscale}

\usepackage{tikz}
\usepackage{tikz-qtree}

\usepackage{latexsym} \usepackage{amssymb}
\title{Counting difficult tree pairs with respect to the rotation distance problem} 
\author{ Sean Cleary and  Roland Maio}
\begin{document}

\maketitle

\begin{abstract}
Rotation distance between rooted binary trees is the minimum number of simple rotations needed to transform one tree into the other.
Computing the rotation distance between a pair of rooted trees can be quickly reduced in cases where there is a common edge between the trees, or where a single rotation introduces a common edge.  Tree pairs which do not have such a reduction are difficult tree pairs, where there is no generally known first step.  Here, we describe efforts to count and estimate the number of such difficult tree pairs, and find that the fraction decreases exponentially fast toward zero.  We also describe how knowing the number of distinct instances of the rotation distance problem is a helpful factor in making the computations more feasible.
 \end{abstract}

\newtheorem{definition}{Definition}
\newtheorem{theorem}{Theorem}
\newtheorem{corollary}{Corollary}
\newtheorem{lemma}{Lemma}
\theoremstyle{plain}
\newtheorem{proposition}{Proposition}
\newcommand{\ds}{\displaystyle}

\section{Introduction}
Trees are a fundamental data structure throughout algorithms and storage implementations.  There are often many different trees representing the same underlying data.  For example, there are many potential trees representing an ordered set of sorted items via a binary search tree.  These different representations may vary considerably in their degree of balance or optimality with respect to a range of criteria.

\begin{figure}
    \centering
       
    \begin{tikzpicture}
\begin{scope} [node/.style={circle,draw},xshift=-0.25cm,scale=0.65,yshift=1.8cm]
   
	\node (r-1) at (0, 0) [node] {$$};
	\node (a-1) at (-1, -1) [node] {$$};
	\node (0-1) at (-2, -2) [node] {$$};
	\node (b-1) at (0, -2) [node] {$M$};
	\node (c-1) at (-1.5, -3) [node] {$$};
	\node (1-1) at (-2.5, -4) [node] {$$};
	\node (2-1) at (-0.5, -4) [node] {$$};
	\draw[-] (c-1) -- (1-1);
	\draw[-] (c-1) -- (2-1);
	\node (d-1) at (1.5, -3) [node] {$N$};
	\node (3-1) at (0.5, -4) [node] {$$};
	\node (e-1) at (2.5, -4) [node] {$$};
	\node (4-1) at (1.5, -5) [node] {$$};
	\node (5-1) at (3.5, -5) [node] {$$};
	\draw[-] (e-1) -- (4-1);
	\draw[-] (e-1) -- (5-1);
	\draw[-] (d-1) -- (3-1);
	\draw[-] (d-1) -- (e-1);
	\draw[-] (b-1) -- (c-1);
	\draw[-] (b-1) -- (d-1);
	\draw[-] (a-1) -- (0-1);
	\draw[-] (a-1) -- (b-1);
	\node (6-1) at (1, -1) [node] {$$};
	\draw[-] (r-1) -- (a-1);
	\draw[-] (r-1) -- (6-1);
\end{scope}
\end{tikzpicture}
\begin{tikzpicture}
\begin{scope} [node/.style={circle,draw},xshift=-0.25cm,scale=0.65,yshift=1.8cm]
  
	\node (r-1) at (0, 0) [node] {$$};
	\node (a-1) at (-1, -1) [node] {$$};
	\node (0-1) at (-2, -2) [node] {$$};
	\node (b-1) at (0, -2) [node] {$N'$};
	\node (c-1) at (-2.5, -4) [node] {$$};
	\node (1-1) at (-3.5, -5) [node] {$$};
	\node (2-1) at (-1.5, -5) [node] {$$};
	\draw[-] (c-1) -- (1-1);
	\draw[-] (c-1) -- (2-1);
	\node (d-1) at (-1.5, -3) [node] {$M'$};
	\node (3-1) at (-0.5, -4) [node] {$$};
	\node (e-1) at (1.5, -3) [node] {$$};
	\node (4-1) at (.5, -4) [node] {$$};
	\node (5-1) at (2.5, -4) [node] {$$};
	\draw[-] (e-1) -- (4-1);
	\draw[-] (e-1) -- (5-1);
	\draw[-] (d-1) -- (3-1); 
	\draw[-] (d-1) -- (c-1);
	\draw[-] (b-1) -- (e-1);
	\draw[-] (b-1) -- (d-1);
	\draw[-] (a-1) -- (0-1);
	\draw[-] (a-1) -- (b-1);
	\node (6-1) at (1, -1) [node] {$$};
	\draw[-] (r-1) -- (a-1);
	\draw[-] (r-1) -- (6-1);
\end{scope}

\end{tikzpicture}
  \caption{An example of a left rotation at node $M$, with a rotation promoting node $N$ to $N'$ and demoting node $M$ to $M'$.}
    \label{figrotation}
\end{figure}
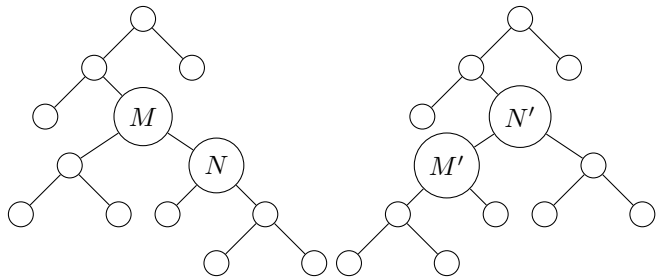

Given a rooted binary tree $S$ where each node has either 0 (``leaf nodes'') or 2 ("interior nodes") children with an order on a set of leaves, we define a {\em left rotation} at an internal node $M$ with right child which is an interior node $N$ as the order-preserving operation which results in a tree where node $M$ is demoted to become a left child of $N$ and $N$ is promoted to take the place of $M$, as shown in Figure \ref{figrotation}.  The former left child of $M$ is now a left-left grandchild of $N$, the former left child of $N$ becomes the right child of $M$, and the former right child of $N$ remains a right child of $N$.  This operation is a local move which does not affect the parent-child relationships of other nodes.  The inverse operation is a {\em right rotation} at node $N'$ which has an interior node as its left child.    Though a single rotation does not change the structure of the tree much, any tree $S$ with a set of ordered leaves can be transformed to any tree $T$ via a sequence of such rotations, as described by Culik and Wood \cite{culik1982note}.

Thus, given two rooted binary trees $S$ and $T$ both respecting an order on an identical set of leaves,  we define the rotation distance $d_R(S,T)$ as the minimum length of a sequence of rotations at nodes to transform  $S$ to $T$.  Rotation distance was introduced and described by Culik and Wood \cite{culik1982note} and work by Sleator, Tarjan, and Thurston \cite{slt88} showed a number of foundational properties of rotation distance, including a sharp upper bound of $2n-6$ for the distance between two trees with $n$ internal nodes for extremely large $n$.  Recent work of Pournin \cite{pournin} gave specific concrete examples realizing these sharp bounds for all $n$ at least 11.

Binary trees are often used for storing ordered data for search applications, and there is a wide range in the efficiency of such operations depending upon the shape of the tree.  Balanced trees result in logarithmic worst-case search time, but stringy trees result in linear worst-case time.  So there has been sustained interested in using local moves such as rotations to balance trees for better expected performance.

There are no known polynomial-time algorithms for computing rotation distance.  There are approximation algorithms which are efficient (see Baril and Pallo \cite{pallo}, Cleary and St. John \cite{cleary2009linear}) and Cleary and St.~John \cite{rotfpt} showed the rotation distance problem to be fixed-parameter tractable.  These algorithms make use of several important conclusions from Sleator, Tarjan, and Thurston about the rotation distance between two trees $S$ and $T$.  First, that if $S$ and $T$ share a common edge, then any minimal length path of rotations does not change that edge and thus the rotation distance problem splits into two smaller subproblems.  If $S'$ and $S''$ are the trees corresponding to the regions on either side of the common edge in $S$, and  $T'$ and $T''$ are the corresponding trees on either side of the common edge in $T$, then $d_R(S,T)= d_R(S',T') + d_R(S'',T'')$.  In this case, we say that the reduction arises from a {\em common edge.}  Second, that if there is an edge in $S$ or $T$ which is one rotation away from becoming a common edge with the relevant other tree, then there is a minimal length path from $S$ to $T$ which either begins or ends with that rotation.  In this case, there is thus a similar reduction giving $d_R(S,T)= d_R(S',T') + d_R(S'',T'')+1$, where $S'$ and $S''$ are the trees corresponding to the regions on either side of the common edge after the single rotation in $S$, with $T'$ and $T''$ corresponding in $T$. In this case, we say that the reduction arises from a {\em one-off edge} and the manner of proceeding is to follow a   {\em one-off move.}

Thus the general rotation distance problem for the distance between two trees $S$ and $T$ can be approached by reducing via common edges and one-off moves, turning the original problem into a collection of smaller subproblems between pairs of trees which do not have any common edges or one-off edges.  These problems form the kernel of the difficulty of the rotation distance problem.

The number of rooted trees of size $n$ is the $n$th Catalan number $C_n$, and the number of pairs of rooted binary trees is $C_n^2$.  So the number of possible instances of the rotation distance problem of size $n$ is $C_n^2$.    We are interested in counting and estimate the number of such instances which are {\em difficult}, meaning that there are no common edges or one-off edges present in the pair.

Cleary, Elder, Rechnitzer and Taback \cite{randomf} calculated the asymptotic density of tree pairs with at least one common edge of a peripheral type, in the context of studying the typical form of elements of the abstract group Thompson's group $F$ (see Cannon, Floyd, and Parry \cite{cfp} or Cleary \cite{ohggt} for the connections between rooted binary trees and Thompson's group.)  Asymptotic combinatorial methods showed that the number of trees pairs without at least one common peripheral edge grows at a rate of $\frac{(8+4\sqrt{3})^n}{n^3} \sim \frac{14.93^n}{n^3}$, and since the number of all tree pairs is simply the square of the Catalan numbers, they grow at the rate of   $\frac{16^n}{n^3}$ and the fraction of tree pairs without such a common edge thus goes asymptotically to zero at an exponential rate of about $( \frac{14.93}{16})^n \sim 0.933^n$.   Since this covers only one possible type of common edge (``peripheral'' in the sense that the relevant tree pair has common subtrees which include all of the leaves of the subtrees) this is a lower bound for the number of trees which have more general common edges.  So it is clear that the fraction of trees with no common edges is smaller than this rate, known to be exponentially converging to 0, and since the fraction of trees with no common edges and no one-off edges is even smaller, that also converges to 0 exponentially fast.  Here, we investigate the rates of these convergences with both some exact and approximate calculations and find estimates for the rates of exponential decay, showing that these phenomena both become exceedingly rare in larger size tree pair examples.  The fraction of tree pairs of size $n$ with no common edges is estimated to be about $0.46 \times 0.916^n $ and the  fraction of tree pairs of size $n$ which are difficult is estimated to be about  $0.094 \times 0.77^n$.

\section{Background}

\begin{figure}[htbp]
  \centering
    \includegraphics[width=0.80\textwidth]{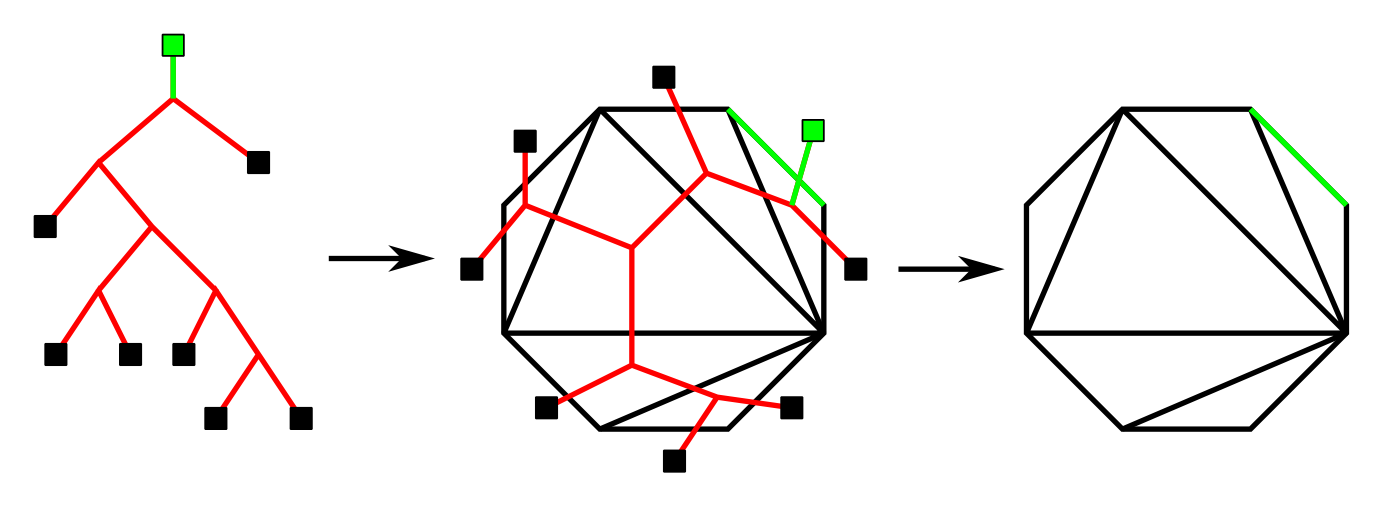}
  \caption{The duality between a rooted binary tree of size $6$ with $7$ leaves into its corresponding triangulation of an octagon with a marked edge corresponding to the root. An exterior edge of the polygon corresponds to either a leaf or the root of the tree, and an interior edge of the triangulation corresponds to an edge in the tree between two interior nodes of the tree.  Figure from \cite{commonedges}.}
  \label{fig:binarytreeconvert}
\end{figure}

By {\em tree of size $n$} we mean a rooted tree with $n+1$ leaves and $n$ internal nodes, where each internal node has exactly two children.  Such trees are also known as $0-2$ trees, see Knuth \cite{knuth3}.  By {\em triangulation of a regular polygon} we mean a collection of non-crossing edges from vertices of a regular $n+2$-gon which divide the polygon into triangles.  There is a natural duality between trees of size $n$ and triangulations of regular $n+2$-gons (see Knuth \cite{knuth3})  illustrated in Figure \ref{fig:binarytreeconvert}.

The notions of {\em right rotation, left rotation} and {\em rotation distance} are described above in Section 1. 
The {\em rotation distance problem}  is the question of given two trees of the same size, to find the rotation distance between the two trees.  

\begin{figure}[htbp]
  \centering
    \includegraphics[width=0.70\textwidth]{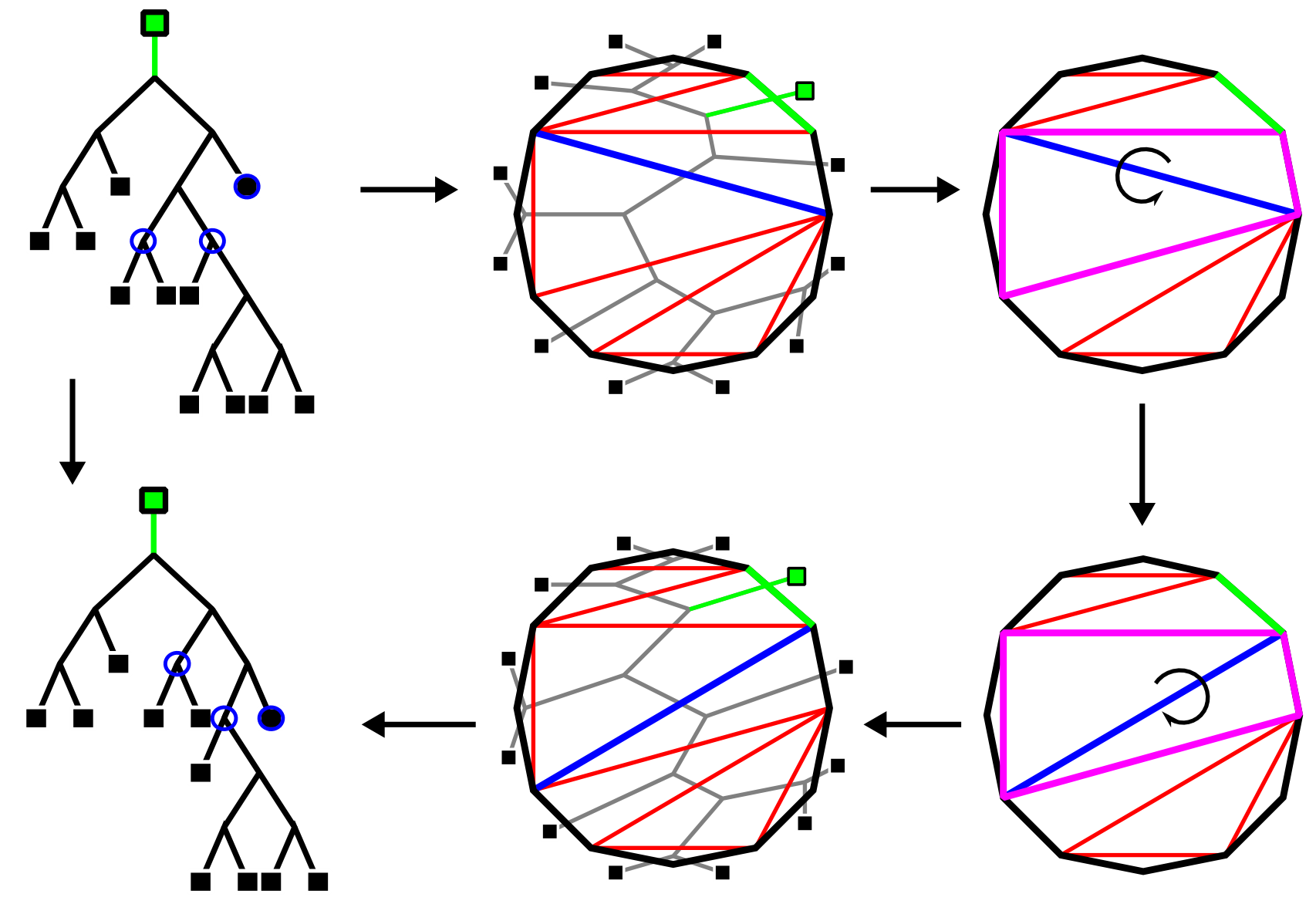}
  \caption{An example of a rotation at a node in a rooted binary tree of size $10$ and the associated edge flip in a dodecagon. The top left tree transforms to the bottom left tree by a right rotation at the right child of the root.  The top right triangulation transforms to the bottom right triangulation by the edge flip across the quadrilateral highlighted in purple, with the transformed edge highlighted in blue.  Figure from \cite{commonedges}.}
  \label{fig:move12}
\end{figure}

Two triangulations differ by an edge flip if they differ by a single change of edge. For example the two rightmost triangulations in Figure   \ref{fig:move12} differ only by the edges marked in blue.  Specifically,
any two adjacent triangles in a triangulation form a quadrilateral, and
an {\em edge flip} of a triangulation of a polygon is the replacement of an edge with the edge between the vertices of the opposite corners of a quadrilateral with a triangle adjacent to the edge of the first triangle.  

The {\em edge-flip distance} between two triangulations of a polygon is the minimum number of edge flips required to transform one triangulation to the other.   Typically, we take the polygon to be convex and in fact regular to have the most symmetric case.  There is a natural duality between finding the edge-flip distance between two triangulations $S'$ and $T'$ of regular polygons of the same size each with a marked edge and finding the rotation distance between their dual rooted trees $S$ and $T$, illustrated in Figure \ref{fig:move12}.

The duality between rotations and edge flips is described in many settings, see  \cite{knuth3, badconflicts, slt88}.
Though the two notions of rotations and edge flips are equivalent through this duality, there are some phenomena which are more easily appreciated from one perspective rather than the other.

The Catalan numbers are $C_n = \frac{(2n)!}{n!(n+1)!} = \frac{1}{n+1} \binom{2n}{n}$ for non-negative integers $n$ and the number of rooted trees of size $n$ (that is, with $n$ interior nodes and $n+1$ leaves) is $C_n$.  Thus the number of rotation distance problems of size $n$ is $C_n^2$, and since the Catalan numbers are asymptotic to $\displaystyle \frac{\displaystyle4^n}{\displaystyle n^{\frac32}\sqrt{\pi}}$ , the number of rotation distance problems of size $n$ grows as $\displaystyle \frac{\displaystyle16^n}{\displaystyle n^3}$ up to a multiplicative constant of $\frac{1}{\pi}$.


\section{Symmetries and classes of rotation distance problems}

From the viewpoint of edge-flip distance between triangulations, it is clear that there are many different instances of what are essentially the same edge-flip distance problem.    These are apparent when we view a triangulation as equivalent to any of its images under a dihedral motion which is a symmetry of the regular polygon.    That is, if we rotate all of the dodecagons in Figure \ref{fig:move12} by $\pi/6$ counterclockwise, the resulting edge-flip move from the rotated dodecagon remains the same.   For a particular instance of a $n+2$-gon edge-flip distance problem, there may be up to $2n+4$ apparently different yet in fact equivalent instances of the same problem.  If the triangulations have some dihedral symmetries, there may not be the full set of $2n+4$ equivalent instances.

There is a long history of counting the number of polygonal triangulations up to dihedral equivalence; see Guy \cite{guy}, Motzkin \cite{motzkin}, or Sloane's sequence A207 in the Online Encyclopedia of Integer Sequences \cite{sloane} for further details.

From the standpoint of computing the number of cases with no common edges or difficult instances, it is only necessary to count the number of such instances up to dihedral equivalence, and then multiply each instance by the number of distinct problems associated to this case.  The number of distinct triangulations up to dihedral equivalence is given by

$$d(n) = \frac{C_n}{2n}  + \frac{C_{n/2+1}}{4} + \frac{C_k}{2} + \frac{C_{n/3+1}}{3} $$

where $k = \frac{n+1}{2}$ for odd $n$, and $k = \frac{n}{2}+1$ for even $n$, and $C_n$ is the $n$th Catalan number for integral $n$ and 0 otherwise, see Sloane \cite{sloane}.   The number of distinct instances of the rotation distance problem is bounded above by $d(n)C_n$ which compares favorably with $C_n^2$ and makes exhaustive computations more feasible, although still exponential in number.


\section{Exhaustive experimental results}

For small size, it is feasible to enumerate all possible instances of the rotation distance problem and tabulate the fractions that have no common edge and no one-off edges.  The results of such enumeration, using the method described above of considering problems up to dihedral equivalence and then counting appropriately, are shown in Table \ref{exacttab}.  Though the calculations used dihedral equivalence to make the calculations more feasible, the results shown below are for all pairs of trees.  An estimate of the rates of growth for the number of cases with no common edges (presuming that there is the same $n^3$ factor as proven in the Catalan and peripheral edge cases) gives growth at about $\displaystyle  \frac{\displaystyle 15.141^n}{\displaystyle  n^3}$ for the number with no common edges, lower than the asymptotic bound of  $\displaystyle \frac{\displaystyle16^n}{\displaystyle n^3}$ for all such instances.

\begin{table}
\begin{center}
\begin{tabular}{|r|r|r|r|} \hline
Size & No common  & Difficult & Total \\ \hline
3 & 10 & 0 &  25 \\
4 & 68 & 8 &  196 \\
5 & 546 & 42 &  1764\\
6 & 4872 & 304 &  17424\\
7 & 46782 & 2616 &  184041\\
8 & 474180 & 23150 &  2044900\\
9 & 5010456 & 209638 &  23639044 \\
10 & 54721224 & 1947692 &  282105616\\
11 & 613912182 & 18501730 &  3455793796 \\
12 & 7042779996 & 179062646 &  43268992144\\
13 & 82329308040 & 1760984370 &  551900410000 \\
14 & 978034001472 & 17561480528 &  7152629313600 \\ \hline
\end{tabular}
\end{center}
\caption{The number of rotation distance problems which have no common edges, are difficult, and the total number of instances $C_n^2$. \label{exacttab}} \end{table}

For small cases, almost all of the common edges are indeed of the peripheral type, so if we instead use a model fitted to the data for sizes 10 through 14 with the presumed $n^3$ factor, we get an exponential growth rate of  about $14.88$, lower than the upper bound of about $14.93$ coming from the peripheral edges only.  As noted in the computational experiments, many of the found common edges are the peripheral type although that fraction decreases as $n$ increases.

\begin{figure}
\includegraphics[width=\textwidth]{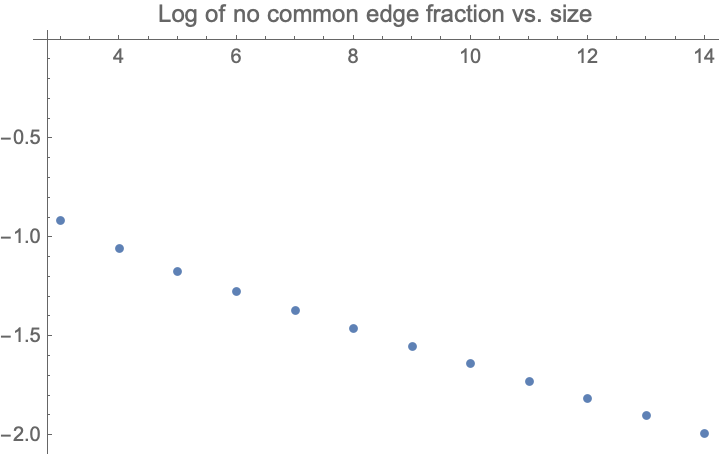}
\caption{Log of the fraction of instances of all tree pairs which have no common edges with respect to size from exhaustive calculation. \label{lognocommexact}}
\end{figure}

\begin{figure}
\includegraphics[width=\textwidth]{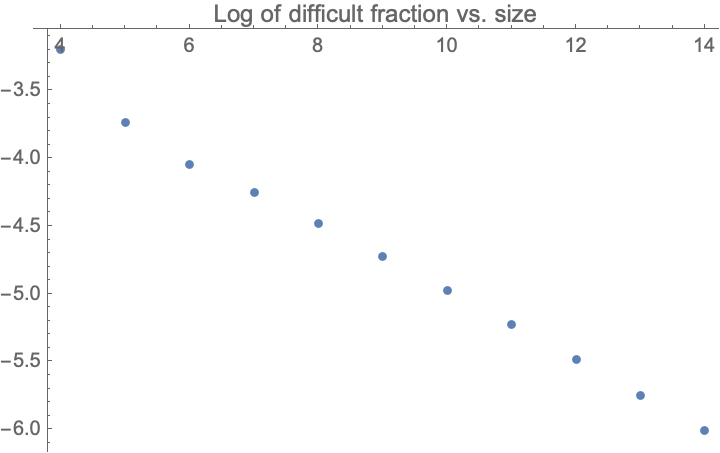}
\caption{Log of the fraction of instances of all tree pairs which are difficult with respect to size from exhaustive calculation. \label{loghardexact}}
\end{figure}

For the number of difficult instances, we again presume there to be a cubic factor of $n$ in the denominator and fit an exponential model of the form $\frac{A^n}{n^3}$ and find a growth rate for $A$ of about $12.5$ depending on how many of the initial terms we ignore as not representative.

If instead of presuming an $n^3$ factor in the denominator, we consider instead the fraction of instances of all instances of the rotation distance problem with no common edges, we get the data shown in Figure \ref{lognocommexact} which has a fitted exponential growth rate of -0.095, giving that the fraction of instances with no common edges shrinking at a rate of about $0.091^n$.  We note that if there is a cubic denominator factor present, this would cancel out in the fraction which agrees with the observed fit well.

For difficult instances as a fraction of all instances, a similar analysis gives Figure \ref{loghardexact} with a fitted growth ratio of about  -0.0257, giving that the fraction of instances with no common edges shrinking at a rate of about $0.77^n$.

We note that indeed these are increasingly sparse, with about 1 in 400 instances of size 14 being hard and about 1 in seven instances having a common edge.  With a decay rate of about 0.77, that gives an estimate of about $4 \times 10^{-13}$ as the fraction of trees of size 100 which are difficult, with only about 1 in 18,000 of size 100 having a common edge of any type.

\section{Results from sampling}

Owing to the exponential growth of the size of the problem,  exhaustive search is not feasible for larger $n$ even taking advantage of the dihedral multiplicity described in the previous section.  Here we present results from sampling tree pairs to estimate the fractions of tree pairs of larger sizes which have no common edges and those which are difficult tree pairs.  The method samples tree pairs uniformly at random using the procedure developed by Remy \cite{remy} for growing a rooted tree of a specified size to produce a tree with uniform probability.  Two such trees of the same size are produced, then the pair is tested for common edges and one-off edges.  Owing to the sparseness of these examples with increasing size, millions of iterations are required to get even one example of a hard pair for sizes 50 and larger.

\begin{table}
\begin{center}
\begin{tabular}{|r|r|r|r|} \hline
Size range & No common  & Difficult & Total sampled in range \\ \hline
15-19  & 3714911 & 48706 & 32m  \\ 
20-24 &1824001 &10142 & 24m \\
 25-29  & 1178295 & 2599 & 24m  \\ 
 30-34  & 2005016 & 1442 & 70m \\ 
 35-39 & 1293851 & 358 & 70m  \\ 
 40-44  & 710261 & 75 & 62m  \\ 
 45-49  & 475554 & 15 & 64m  \\ 
 50-54  & 729537 & 18 & 125m  \\ 
 55-64  & 601860 & 7 & 160m  \\ 
 65-74  & 372756 & 1 & 240m  \\ 
 75-84  & 77034 & 0 & 120m  \\ 
 85-94  & 34205 & 0 & 130m  \\ 
 95-104  & 1113 & 0 & 10m  \\ 
 105-114  & 918 & 0 & 18m  \\ 
 115-124  & 380 & 0 & 18m  \\ 
 125-134  & 143 & 0 & 18m  \\ 
 135-144  & 24 & 0 & 12m  \\ 
 155-164  & 6 & 0 & 12m  \\ 
 165-184  & 2 & 0 & 12m  \\ 
 195-210  & 0 & 0 & 12m\\ \hline
\end{tabular}
\end{center}
\caption{The number of observed rotation distance problems which have no common edges, are difficult, and the number sampled in millions. \label{sampletab}} \end{table}

Results of these sampling experiments are summarized in Table \ref{sampletab} and reflect a large number of runs of sizes of tree pairs from 15 to 207, showing the number of instances of pairs with no common edges, difficult pairs, and the total number of trees pairs test for various size ranges.

\begin{figure}
\includegraphics[width=\textwidth]{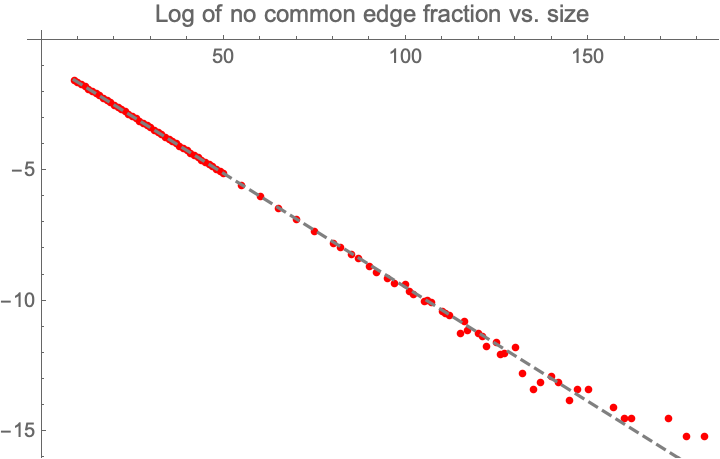}
\caption{Sampled estimates of the log of the fraction of instances of all tree pairs have no common edges with respect to size.  Five sample points with no observed common edges in the range 152 to 170 are omitted from the plot as those have undefined logarithms. \label{samplednocomm}}
\end{figure}

Again the fraction appears to diminish exponentially and we can fit linear models to the log of the number present versus the size considered.  The data for the number of tree pairs with no common edges is shown in Figure \ref{samplednocomm}.  There, we find that the fraction of tree pairs with no common edges is fitted well with an exponential model of  $p_{nocommon} \sim 0.4644 \times 0.91641^n$.

\begin{figure}
\includegraphics[width=0.99\textwidth]{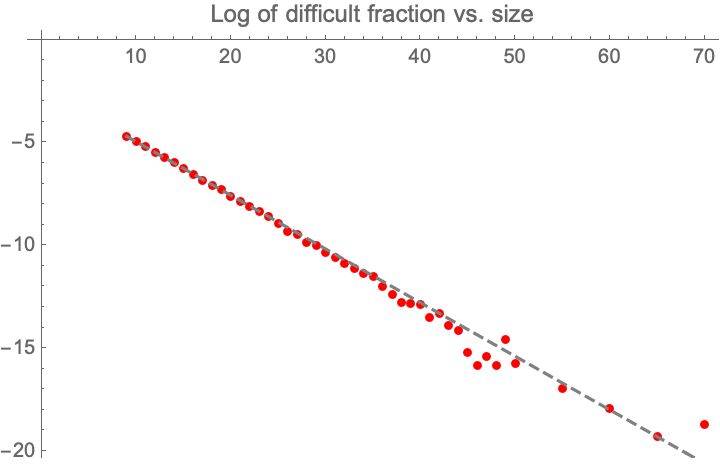}
\caption{Sampled estimates of the log of the fraction of instances of all tree pairs are difficult with respect to size. \label{sampledhard} Size 70 was the largest where there was at least one observed sampled instance of a difficult tree pair.}
\end{figure}

In the case of the difficult instances, we see of course smaller fractions.    The data for the number of tree pairs found to be difficult is shown in Figure \ref{sampledhard}.  Again, an exponential model fits well to data in the range where there are non-zero observations of difficult pairs with a model $p_{hard} \sim 0.09407 \times 0.7705^n$.

\section{Discussion}

Work of Cleary, Elder, Rechnitzer, and Taback \cite{randomf} gave proven asymptotic results for the fraction of tree pairs with no common edges of peripheral type, decreasing exponentially at a rate of about $0.933^n$.  So it is no surprise that the smaller fraction of tree pairs with no common edges of any type appears to decrease more rapidly, at a rate of about $0.91^n$.
The fraction of tree pairs with not only no common edges but further no one-off edges appears also to decrease at an exponential rate of about $0.77^n$.  So these phenomena will become vanishingly rare for large $n$.

That these phenomena are rare is good news from the standpoint of solving the rotation distance problem, as it means that a typical instance of a rotation distance problem of large size is nearly certain to have a common edge or a one-off edge present, allowing some progress forward.  However, from the perspective which wants to understand the general rotation distance problem, it is not easy to find examples which possess the central difficulty of the problem as these become increasingly rare.

One method for sampling difficult pairs would be to sample large tree pairs, then reduce along common edges and one-off edges until a collection of smaller difficult tree pairs arise.  This parallels the description in Cleary, Rechnitzer, and Wong \cite{commonedges} where the questions of what is the expected size of the largest remaining piece after splitting along peripheral common edges is studied and approximated.  This is a method for generating difficult instances of the problem for study, but since the number of reductions and their locations varies, it does not produce a difficult instance of a specified size.   Typically, given a large pair, there would be about  9\% common edges, and the size of the largest remaining component  after reduction is about $(4-\pi)n \sim 0.85 n$.  Experimental work of Chu and Cleary \cite{conflicts} suggested that the number of one-off edges between randomly selected trees was comparable to the number of common edges, as confirmed by the work of  Cleary, Rechnitzer, and Wong \cite{commonedges} showing the number of one-off edges in a piece of size $n$ with no common edges  to be asymptotically $(\frac54-\frac{1}{(4-\pi)})n \sim 0.085n$.  So for example, selecting a tree pair of size 100 at random may result in the largest piece with no coming edges being of size 85, which itself may have about 6 one-off edges, giving perhaps a largest remaining piece of size 63 which corresponds to a difficult pair.   Another random sample may end up with something of size 71 or 58 and so on, resulting in collections of difficult instances across a range of sizes.  So though the vast majority of instances have reductions to smaller problems, the largest of the small problems may still pose significant difficulty.

Cleary and Maio \cite{hardcases} have a heuristic sampling algorithm for constructing difficult instances of the rotation distance problem of a specified size by a growing method based upon the method of Remy \cite{remy} modified to work with tree pairs and which runs in polynomial time.  This is preferable to a ``test-and-reject" approach particularly when interested in difficult examples of a particular size, as the fractions of success quickly diminish and become quite small for increasing $n$.

The authors are grateful for funding provided by NSF award \#1417820. This work was partially supported by a grant from the Simons Foundation (\#234548 to Sean Cleary).

Sean Cleary, Department of Mathematics, The City College of New York, Convent Ave. at 138th St, New York NY 10031, USA. {\tt cleary@sci.ccny.cuny.edu}

Roland Maio, Department of Computer Science, Columbia University, 500 W 120th St., New York, NY 10027, USA.  {\tt roland@cs.columbia.edu}

\end{document}